\documentclass[a4paper,11pt]{article}
\usepackage{jheppub_deleted}
\usepackage{amsfonts}
\usepackage{amsmath,amssymb,amscd}
\usepackage{appendix,marginnote,tikz,pgf,mathtools}
 \usepackage{comment}

\usepackage[numbers]{natbib}
\def\cL{\mathcal{L}}

\def\ll{\left\langle}
\def\rr{\right\rangle}

\def\Tr{{\rm Tr}\,}

\def\be{\begin{equation}}
\def\ee{\end{equation}}  
\def\bea#1\eea{\begin{align}#1\end{align}} 

\def\pd{\partial}
\def\d{\partial}
\def\p{\partial}
\def\t{\tau}
\def\F{{\cal F}}
\def\ll{\left\langle}
\def\rr{\right\rangle}

\makeatletter

    \@addtoreset{equation}{section}
\makeatother

\title{From minimal gravity to open intersection theory }
\author[a,b]{Alexander Alexandrov}
\author[c]{, Hisayoshi Muraki}
\author[c]{and Chaiho Rim}
\affiliation[a]{Center for Geometry and Physics, Institute for Basic Science (IBS), Pohang 37673, Korea}
\affiliation[b]{ITEP, Moscow, Russia}
\affiliation[c]{Department of Physics, Sogang University, Seoul 04107, Korea}

\emailAdd{alexandrovsash@gmail.com}
\emailAdd{hmuraki@sogang.ac.kr}
\emailAdd{rimpine@sogang.ac.kr}

\abstract{ 
We investigated the relation between the two-dimensional minimal gravity 
(Lee-Yang series) with boundaries and  open intersection theory. 
It is noted that the minimal gravity with boundaries is 
defined in terms of boundary cosmological constant $\mu_B$
and the open intersection theory in terms of boundary marked point generating parameter $s$.
Based on the conjecture that the two different descriptions of the 
generating functions are related by the Laplace transform,
we derive the compact expressions for the generating function of the intersection 
theory from that of the minimal gravity on a disk and on a cylinder. 
}

\begin{document} 
\maketitle
\flushbottom

%%%%%%%%%%%%%%%%%%%%%%%%%%%%%%%%%%%%%%%%%%%%%%%%%%%
\section{Introduction} 
The interplay of the KdV hierarchy between
two-dimensional quantum gravity and intersection theory (IT)
for the moduli spaces of closed Riemann surfaces
was proposed in \cite{Witten_91}. 
The conjecture of Witten for the IT  was proved in \cite{Konst_92}
by finding the generating function (GF), also known as free energy,
of the one-matrix Airy function.
Later, the KdV hierarchy 
was also checked for 2d minimal gravity of Lee-Yang series (MG)
by constructing  the effective GF 
on genera up to 3 in \cite{BBT_11}
using the one-matrix model polynomials.

Similar interplay is noted for the models on the Riemann surfaces 
with boundaries \cite{BMR_18,MR_18}.
The GF of the MG with boundaries (BMG) 
satisfies the relations, similar to the ones known for the open intersection theory (OIT)
 \cite{PST_14}. 
These relations include
the boundary version of  the KdV hierarchy 
and the string equation,
which encode the generalized Virasoro constraints, 
the analogue of the constraints 
on the closed Riemann surfaces 
\cite{FKN_91,  DVV_91}.

In this paper we continue the investigation of the interplay between the two-dimensional gravity 
and intersection theory on the moduli spaces of Riemann surfaces with boundaries.  

While the GF of IT and that of MG satisfy essentially the same equations, 
including the KdV hierarchy and the Virasoro constraints, they have
perturbative expansions with drastically different properties. 
The reason is that they belong to different classes of the solutions with different types of analytical properties, 
which can be related with each other by different types of analytical continuation. 
It should be noted that 
the set of KdV parameters in two theories 
plays a different role. 
On the closed Riemann surfaces, 
the bulk cosmological constant $\mu$ provides 
the gravitational scaling dimension (GSD)  and 
plays a major role for MG.
In addition the GF is non-analytic in $\mu$
and therefore, 
one cannot turn off $\mu$ in MG. 
On the other hand, $t_0$ in IT provides the scaling dimension (SD), 
which, in general, has nothing to do with $\mu$.  
As a result, even though both GF are solutions of the 
same KdV hierarchy, they are different.

Similar analysis holds for the open KdV hierarchy. 
BMG needs the boundary cosmological constant $\mu_B$
and its open KdV hierarchy is given in terms of  $\mu_B$.
On the other hand, OIT and its open KdV are described by the boundary parameter $s$.
As a result, the open KdV hierarchy of BMG differs
from that of OIT. Nevertheless, the two different KdV hierarchies turn out to be closely related each other: 
Exponentials of the GF of two theories
are related through the Laplace transform \cite{MR_18}.

As a result of the comparison between two representations, we obtain the 
elegant formulas for the generating functions on the disk and the cylinder. They are given in terms of the 
variables, which are known to be convenient for the description of the matrix models \cite{AJM,ACKM}. 

The paper is organized as follows. 
In  section \ref{S2}, description of  IT and MG in terms of the KdV hierarchy and Virasoro constraints are summarized and clarified:
Section \ref{S2.1} is for the closed  Riemann surfaces  
and section \ref{S2.2} is for the Riemann surfaces with boundaries
(open Riemann surfaces). In section \ref{S3}, GF of BMG with $\mu_B$-parameter
is analyzed in terms of open KdV hierarchy 
with Euler characteristic expansion.
In section \ref{S4}, we find 
GF of OIT in $s$-space 
using the Laplace transform
from GF of BMG in $\mu_B$-space.
In this process, one has to find a way to
choose the right solution of string equation. 
We check explicitly that 
the transformed GF in $s$-space 
coincides with the known GF of OIT on a fluctuating disk.
In addition we provide the GF on a fluctuating cylinder.
Section \ref{S5} is the summary and discussion.

%%%%%%%%%%%%%%%%%%%%%%%%%%%%%%%%%%%%%%%%%%%%%%%%%%%%
\section{Minimal gravity vs intersection theory}\label{S2}

%%%%%%%%%%%%%%%%%%%%%%%%%
\subsection{Riemann surfaces without boundary}\label{S2.1} 
%%%%%%%%%%%%%%%%%%%%%%%%%%

Minimal quantum gravity of Lee-Yang series $M(2, 2p+1)$ 
on the closed Riemann surface is  described  
either by the Liouville field theory coupled to conformal matter 
 \cite{KPZ_88} 
or by the scaling limit of one-matrix model \cite{KKM_85, K_89}. 
Its generating function (free energy) has natural genus expansion 
\be
\label{genus-expansion-MG}
{\cal F}^c 
=\sum_{g=0}^\infty \lambda^{2g-2}{\cal F}^c _{(g)},
\ee
where $\lambda$ is the genus expansion parameter. Two approaches led to the different expressions for the GF, 
which are believed to be related to each other by the so-called resonance transformations \cite{MSS_91,BZ_09}.
Below we will work  with the matrix model description, which has a much clearer relation to the integrable hierarchies. 
The GF ${\cal F}^c$ depends on the multiple KdV parameters  $\t_0, \t_1, \cdots, \t_{p-1}$, 
and it is convenient to introduce the function
\be
\label{u}
u\equiv \frac{\pd^2 {\cal F}^c }{\pd \t^2_0}.
\ee 
The matrix model formulation allows us to construct the GF, ${\cal F}^c$, dependent on the infinitely many descendent variables. However, we do not consider this opportunity below.

The flow equations of $u$ 
along the KdV parameter directions
constitute the KdV hierarchy  \cite{Doug_90, BDSS_90,GM_90,GGPZ_90}.
The KdV hierarchy and the string equation satisfied by the GF of MG can be represented as 
 \begin{align}
 \label{MG-hierarchy}
\frac{1}{\lambda^{2}}\frac{2n+1}{2}\frac{\pd^3 {\cal F}^c }{\pd \t^2_0\pd \t_n}
&=\frac{\pd^2{\cal F}^c }{\pd \t^2_0}\frac{\pd^3 {\cal F}^c }{\pd \t^2_0\pd \t_{n-1}}
+\frac{1}{2}\frac{\pd^3 {\cal F}^c }{\pd \t^3_0}\frac{\pd^2 {\cal F}^c }{\pd \t_0\pd \t_{n-1}}
+\frac{1}{8}\frac{\pd^5 {\cal F}^c }{\pd \t^4_0\pd \t_{n-1}},
\\
\label{MG-string}
0&=\sum_{n\geq0}\t_{n+1}\frac{\pd {\cal F}^c }{\pd \t_n}+\frac{\t^2_0}{2\lambda^2},
\end{align} 
where $\tau_{p+1}=1$ and $\tau_k=0$ for $k>p+1$.

On a fluctuating sphere, the flows are described by the dispersionless limit of the KdV hierarchy \eqref{MG-hierarchy}, 
which has the simple form
\be
\label{MG-hierachy-g=0}
	\frac{\partial^3 {\cal F}^c_{(0)}}{\partial \t_n \partial \t_0^2} 
	=\frac{\pd v}{\pd \t_n}
	=\frac{v^n}{n!}\frac{\pd v}{\pd \t_0}
	\qquad{\rm for}\qquad
	1 \le n\le p-1.
\ee 
Here  ${\cal F}^c_{(0)}$ is the GF on the sphere,
 and it is best described in terms of $A_1$ Frobenius manifold 
\cite{BDM_14,Dub_92}
(see also \cite{BR_15, ABR_17} for a dual description in terms of $A_{2p}$)
whose coordinate is identified with the second derivative of the GF: 
\be
	v\equiv \frac{\pd^2 {\cal F}^c_{(0)} }{\pd \t^2_0}.
\ee

The string equation can be reduced to the polynomial form:
\be 
\label{string_polynomial_2d_gravity} 
	{\cal P}(\t, v)=0;
	\qquad  
	{\cal P}(\t, v) \equiv \sum_{m=0}^{p-1} \t_m  \frac{v^{m}}{m!} +  \frac{v^{p+1}}{(p+1)!}.
\ee 
This equation can be obtained  if one takes the second derivative of \eqref{MG-string} with respect to $\t_0$, 
uses the dispersionlees KdV hierarchy \eqref{MG-hierachy-g=0} 
and integrates the result over $\tau_0$. 

It is noted that GF of MG is constructed on the fluctuating sphere
in \cite{Z_05} for the Lee-Yang model using both of the results of Liouville field theory and matrix model.
This result is extended  in \cite{BZ_09} to the Lee-Yang series
\be 
\label{MG_GF_sphere}
	{\cal F}^c_{(0)} = \frac 12 \int^w_0  {\cal P}^2(v)  dv, 
\ee
where  ${\cal P}(v)$ is the string polynomial and 
$w$ is a proper solution of the string equation of the polynomial 
form \eqref{string_polynomial_2d_gravity}. 
One can check easily that ${\cal F}^c_{(0)}$
satisfies the KdV hierarchy and the string equation.
GF of MG is further constructed up to genus 3 in  \cite{BBT_11}, 
and these contributions are also found to satisfy the KdV hierarchy.

In \cite{Witten_91} it was conjectured that two-dimensional gravity 
is related to the intersection theory on the moduli space of Riemann surfaces. 
The GF of  IT depends on the infinitely many parameters, 
$t=(t_0, t_1, \cdots ) $, playing the role of the coupling constants of the gravitational descendants in topological gravity. 
From this physical identification of different two-dimensional gravity models 
Witten concluded that the all genera GF of IT,  $F^c$, also satisfies the KdV hierarchy and the string equation:
\begin{align}
\label{IT-KdV}
\frac{1}{\lambda^{2}}\frac{2n+1}{2}\frac{\pd^3 F^c}{\pd t^2_0\pd t_n}
&=\frac{\pd^2 F^c}{\pd  t^2_0}\frac{\pd^3 F^c}{\pd t^2_0\pd  t_{n-1}}
+\frac{1}{2}\frac{\pd^3 F^c}{\pd t^3_0}\frac{\pd^2 F^c}{\pd t_0\pd t_{n-1}}
+\frac{1}{8}\frac{\pd^5 F^c}{\pd t^4_0\pd t_{n-1}}, 
\\
\label{IT-string}
\frac{\pd F^c}{\pd t_0}&=\sum_{n\geq0}t_{n+1}\frac{\pd F^c}{\pd t_n}+\frac{t^2_0}{2\lambda^2}\,.
\end{align}
Here and from now on, 
we distinguish the notation of GF,
$\cal F$ for MG and $F$ for IT. 
The conjecture was proved  in  \cite{Konst_92}
by identifying of the GF of IT with the matrix integral over $N\times N$ hermitian matrix $X$ with the cubic potential
\be
e^{F^c}\propto \int \left[d X\right] e^{-\frac{1}{\lambda}\Tr\left(\frac{X^3}6 -\frac{\Lambda^2 X}2 \right)},
\ee
with the condition  $F^c(t=0) =0$.
For this Kontsevich matrix integral representation the set of KdV parameters
is given by the Miwa variables
\be
t_k=\lambda\, (2k-1)!!\, \Tr \Lambda^{-2k-1},
\ee
assuming $N$ is sufficiently large.

It is known that 
the KdV hierarchy and string equation imply 
the Virasoro constraints  \cite{FKN_91,DVV_91}:
\be\label{closed_Virasoro_constraint}
L_n e^{F^c} =0 \quad{\rm for}\quad n \ge -1. 
\ee 
The Virasoro generators are 
\begin{align}
	L_{-1}  
	&=\sum_{i\ge0} (t_i - \delta_{i,1}) \frac{\partial }{\partial t_{i-1}} + \frac{t_0^2}{2 \lambda^2},\\
	L_{0} 
	&=\sum_{i\ge0} \frac{2i +1}2 (t_i - \delta_{i,1}) \frac{\partial }{\partial t_{i}} + \frac{1}{16},
\end{align}
and for $n>0$ 
\begin{align}
	L_n  
	&=\sum_{i\ge0}\frac{(2i +2n +1)!! }{2^{n+1} (2n-1)!! }(t_i - \delta_{i,1}) \frac{\partial }{\partial t_{i+n}} 
	\nonumber\\
	&\qquad\qquad
	+\frac{\lambda^2} 2 \sum_{i \ge 0}^{n-1} \frac{(2i+1)!!(2n-2i -1)!!}{2^{n+1}}\frac{\partial^2}{\partial t_i \partial t_{n-i-1}}.
\end{align}
The Virasoro constraints can also be derived from the Kontsevich matrix integral \cite{W_91,MMM,KMMM}.

 From the comparison of the KdV hierarchy and 
the string equation of MG \eqref{MG-hierarchy}-\eqref{MG-string}  and 
those of IT \eqref{IT-KdV}-\eqref{IT-string}  one can conclude, that they coincide  after an identification of all variables $\tau_n$ with  $t_n$ except for $n=1$. Namely,
$t_1$  is identified with $\t_1$  shifted by a constant,
$t_1=\t_1 +1$. This is a well-known dilaton shift. 
Let us stress that, while the equation satisfied by the GF of IT and
that of  MG are  almost the same, the solutions do not coincide. In particular, the role of KdV parameters in two models differs dramatically. 
First, $\t_n$ in MG couples to the gravitation primary operator of Lee-Yang series.
On the other hand,  $t_n$ for $n>0$ in IT couples
to gravitational descendant operator.  

Second, more important is the role of $\t_{p-1}$ of MG and $t_0$ of IT. 
In MG approach the cosmological constant $\mu$ 
should be present 
and all the physical quantities in MG 
are equipped with the gravitational scaling dimension (GSD),
which counts the power of $\mu$ \cite{KPZ_88}. 
It is known  \cite{MSS_91,Z_05} that $\t_{p-1}$ plays the role of $\mu$.
Other KdV parameters $\t_n$ ($n< p-1$)
are considered as deformation parameters.
Before the deformation, 
the string equation of the polynomial form
\eqref{string_polynomial_2d_gravity},  
\be
\label{MG-string-solution}
 \t_{p-1} \frac{v^{p-1}}{(p-1)!} +  \frac{v^{p+1}}{(p+1)!} =0 ,
\ee
has a non-trivial solution $v \propto \sqrt{\mu}$ and namely this solution describes MG.
This shows that, in general, GF of MG is  non-analytic in $\t_{p-1}$, 
as GF is given in powers of $v$ (see for example, \eqref{MG_GF_sphere})
and GSD is given by a fractional number.

It can be shown that the GF of MG is scale-free.
Note that in  \eqref{string_polynomial_2d_gravity} the coefficient of the term  $v^{p+1}$  is assumed to be
scale-free and can be normalized to 1.
Since GSD of $v$ is 1/2, 
GSD of ${\cal P}(\t, v)$ is $(p+1)/2$
and GSD of the deformation parameters $\t_k$ 
is $(p+1-k)/2$.
According to  \eqref{MG_GF_sphere}, 
GSD of  GF on sphere
is $(2p+3)/2$. 
In addition, the genus expansion parameter $\lambda^2$
in \eqref{genus-expansion-MG} 
has non-vanishing GSD, 
namely, $(2p+3)/2$.

To compare the IT with the MG we can put $t_n=0$ ($n\geq p$) in the GF, which restricts to the subspace with the finite number of KdV parameters 
$t=(t_0, t_1, \cdots t_{p-1}) $. Then, the string polynomial for IT is obtained from \eqref{string_polynomial_2d_gravity} by the above described between $\tau$ and $t$,
\be
\label{IT-string-polynomial}
 P (t, v) 
=  \sum_{m=0}^{p-1} t_m  \frac{v^{m}}{m!}  -v ,
\ee 
where the linear power of $v$ is added due to the $t_1$ shift. In this case, $t_0$ becomes the 
basic parameter and the others
are treated as deformation parameters, so that the undeformed solution is $v = t_0$.  
One can show that the string polynomial 
\eqref{IT-string-polynomial}
is consistent with the  KdV hierarchy  \eqref{IT-KdV} 
and the string equation \eqref{IT-string}
to the lowest order in $\lambda$.

The solution of this string equation,
corresponding to IT is completely perturbative in $t_k$'s.
Namely, it is a power series of all the KdV parameters $t_k$
so it has a regular limit when all of them, including $t_0$, go to zero.
In a certain sense, the GF of IT can be considered as a ``universal'' GF for the whole Lee-Yang series,
starting from $M(2, 1)$ model, adding proper number of  variables and allowing analytic continuation on the solution space \cite{BBT_11}. 

Like in MG, one can introduce the scaling dimension (SD) to IT:
In \eqref{IT-string-polynomial},  
$t_1$ is assumed to be a scale-free parameter. 
Therefore, it is natural to define SD as the power of $t_0$, 
the basic scale parameter.  
This shows that SD of $v$ is 1 and 
therefore, SD of $P(t, v)$  is assigned to be  1
and SD of the deformation parameters $t_k$  is   $1-k$. 
Note that $v=t_0$ before deformation, 
and that SD of GF on sphere is 3 
(from the definition $v= {\d^2 F^c_{(0)}}/{\d t_0^2}$),
so is the  SD of the genus expansion parameter $\lambda^2$.

%%%%%%%%%%%%%%%%%%%%%%%%%
\subsection{Riemann surfaces with boundaries} \label{S2.2}
%%%%%%%%%%%%%%%%%%%%%%%%%%

Recently the intersection theory on the moduli spaces of the Riemann surfaces with boundaries was developed by Pandharipande, Solomon and Tessler \cite{PST_14}, see also \cite{T_15,BT_17}. 
They have described the GF  $F^o$ for the open intersection numbers 
\be\label{OIT-in}
\int_{\overline{\mathcal M}_{\bar{g},k,l}} \psi_1^{a_1} \psi_2^{a_2}\dots \psi_l^{a_l},
\ee
given by the integrals of the products of the first Chern classes 
$\psi_i$  of the cotangent line bundles over the compactification 
${\overline{\mathcal M}_{\bar{g},k,l}} $ of the
moduli space of  Riemann surfaces with boundaries. They also constructed explicitly the leading contribution to this GF, given by the disk geometry, and explained how to make the naive description \eqref{OIT-in} precise for the higher geometries.
The integral is non-vanishing 
only if dimension of  ${\overline{\mathcal M}_{\bar{g},k,l}} $,
 \be
 	\dim_{\mathbb R}{\mathcal M_{\bar{g},k,l}}=3\bar{g}-3+k+2l,
 \ee
coincides with the degree of the integrand:
\be\label{OIT-def}
3\bar{g}-3+k+2l=\sum_{j=1}^l 2 a_j,
\ee
and the stability condition $2\bar{g}-2+k+2l>0$ is satisfied.

The GF, which depends on the KdV parameters $t_k$ and an additional parameter $s$, associated with the insertion of the marked points on the boundary, has a natural topological expansion
\be
\label{OIT-GF-g-expansion}
F^o  
=\sum_{\bar g =0}^\infty \lambda^{\bar g-1}F^o_{(\bar g)}.
\ee
Here $\bar g$ is the genus of the doubled Riemann surface. 
This expansion can be interpreted as the Euler characteristic expansion:
\be 
F^o  
=\sum_{ \chi \le 1 } \lambda^{-\chi}F^o_{(\chi)} ,
\ee
where $\chi= 2 - 2 g -k $ 
is given in terms of the number of handles ($g \ge 0$)
and the number of boundaries ($k \ge 1$) of the Riemann surface with boundaries, and is related to $\bar g$ as $\bar g =1-\chi$.
Hereafter,  we call the Euler characteristics expansion 
 the $\bar g$-expansion.
 
The authors of \cite{PST_14} also suggested a generalization of the Virasoro constraints (\ref{closed_Virasoro_constraint}) for the open case:
\be
\label{open_Virasoro_constraint}
B_n e^{F^c + F^o}=0~~~
{\rm for}~ n \ge -1, 
\ee
where
\be
\label{open_Virasoro_generator}
B_n = L_n + \lambda^n s 
\frac{\partial^{n+1}}{\partial s^{n+1}} 
+ \frac{3n+3}4 \lambda^n 
\frac{\partial^n}{\partial s^n}.
\ee 
In particular, for $n=-1$ with the help of the string equation \eqref{IT-string}, the equation \eqref{open_Virasoro_constraint} reduces to the open string equation 
\be 
\label{OIT-string}
\frac{\partial F^o}{\partial t_0} 
= \sum_{n \ge 0} t_{n+1} \frac{\partial F^o}{\partial t_n}
+\frac{s}{ \lambda}\,.
\ee  

An open version of the KdV hierarchy, satisfied by the open GF, was also introduced in \cite{PST_14}
\be
\label{open_KdV}
\frac{2n+1}{2}\frac{\partial F^o}{\partial t_n}
 =\lambda\frac{\partial F^o}{\partial s}
\frac{\partial F^o}{\partial t_{n-1}}
 +\lambda\frac{\partial^2 F^o}
{\partial s\partial t_{n-1}}
 +\frac{\lambda^2}{2}
\frac{\partial F^o}{\partial t_0}
\frac{\partial^2 F^c}{\partial t_0\partial t_{n-1}}
 -\frac{\lambda^2}{4}
\frac{\partial^3 F^c}{\partial t^2_0\partial t_{n-1}}.
\ee
While the relation of the open KdV equations to the integrable hierarchies remains unclear, it was proven 
by Buryak in \cite{B_15},
that the open KdV hierarchy has a unique solution with the given initial conditions, and this solution satisfies the open Virasoro constraints (\ref{open_Virasoro_constraint}).
Buryak also found an additional $s$-flow equation, 
which is consistent with the open KdV hierarchy:
\be
\label{OIT-s-flow}
\frac{\partial F^o}{\partial s} 
= \lambda \left\{
\frac{1}{2}\left(\frac{\partial F^o}{\partial t_0}\right)^2 
+ \frac12 \frac{\partial^2 F^o}{\partial t_0^2}
+ \frac{\partial^2 F^c}{\partial t_0^2}
\right \}\,.
\ee

Having in mind the connection between IT and MG in the closed case, outlined in section 2.1, 
it is natural to expect a similar connection for the case with boundaries.  The GF of BMG
has the $\bar g$-expansion  \eqref{OIT-GF-g-expansion}:
\be 
{\cal F}^o 
=\sum_{\bar g=0}^\infty \lambda^{\bar g-1}{\cal F}^o_{(\bar g)}.
\ee
However, the description of the OIT looks completely different from that of the BMG. 
Namely, boundary effects in BMG are described by 
the boundary cosmological constant $\mu_B$, whose nature essentially differs from that of the boundary  marked point insertion parameter $s$ of the OIT.  

A clue to the relation between two pictures can be seen from the comparison of the equations, satisfied by the leading terms of the $\bar g$ expansion, that is GF's on the disk. For the BMG on the disk, ${\cal F}^o_{(0)}$, one has
\cite{MSS_91}
\be 
\label{GF_disk}
{\cal F}^o_{(0)} (\tau, \mu_B)
	=\frac {i}{\sqrt{2 \pi}}\int_0^\infty \frac{dl}{l^{3/2}} e^{-l\mu_B}
		\int_{\tau_{0}}^\infty dx \ e^{-l v(x)}\,.
\ee 
Here  $v(x)$ is a function of $x$, which
is governed by the string equation  of the polynomial form \eqref{string_polynomial_2d_gravity}, 
with $\tau_0$ substituted by $x$. 
The GF in \eqref{GF_disk}
satisfies the following equation \cite{BMR_18}:  
 \be
\label{BMG_open_KdV_0}
  \frac {2n+1}2  
\frac{\partial {\cal F}^o_{(0)}}{\partial \t_n}
 = -\mu_B \frac{\partial{\cal F}^o_{(0)}}{\partial \t_{n-1}}+
 \frac{1}{2}\frac{\partial {\cal F}^o_{(0)}}{\partial \t_0}\frac{\partial^2 {\cal F}^c_{(0)}}{\partial \t_0\partial \t_{n-1}},
\ee
for $1\leq n \leq p-1$.
This equation can be easily derived if one notes that
the multiplication  of the integrand of \eqref{GF_disk} by $-\mu_B$
can be replaced by a derivative with respect to $l$.
Details are given in appendix A. This functional relation is similar 
to the open KdV   \eqref{open_KdV} at $\bar g =0$:
${\cal F}^o_{(0)}$ corresponds to $F^o_{(0)}$,
and ${\partial F^o_{(0)}}/{\partial s}$ is  replaced 
by $-\mu_B$, which is assumed to be independent of $\t_n$.

This observation allows us to conjecture, that the $s$ and $\mu_B$ pictures are related by the Laplace (or Fourier) transform \cite{MR_18}
\be
\label{Laplace}
	e^{{\cal F}^o( s)} = 
	\frac{1}{\sqrt{2 \pi \lambda}}\, \int d \mu_B \, e^{-\frac{s\mu_B}{\lambda} } ~ e^{{\cal F}^o(\mu_B)},
\ee
with the inverse transform
\be\label{invLaplace}
	e^{{\cal F}^o( \mu_B)} = 
	\frac{1}{\sqrt{2 \pi \lambda}}\, \int d s \, e^{\frac{s\mu_B}{\lambda} } ~ e^{{\cal F}^o(s)}.
\ee
To avoid confusion we indicate explicitly the variable $s$ or $\mu_B$. 
Here we continue to use the notation ${\cal F}$ after the Laplace transform and, depending on the context, we assume it to depend on the KdV variables $\tau$ or $t$. The reason is that we expect this relation still to be valid beyond the proper parameter range of the OIT of \cite{PST_14}. 
As we will show in the following section, to get the 
GF of OIT one has to choose very particular 
${\cal F}^o(\mu_B)$, which corresponds to 
a certain  topological branch of BMG.

The Laplace transform in \eqref{Laplace} and \eqref{invLaplace} can be interpreted with the saddle point method for the large values of $\lambda$. 
Then the critical values of $\mu_B$ and $s$ in \eqref{Laplace} and \eqref{invLaplace} satisfy respectively
\be
s=\frac{\p{\cal F}^o_{(0)} (\mu_B)}{\p \mu_B},
\ee
and
\be
\label{eq:2.35}
\mu_B=-\frac{\p{\cal F}^o_{(0)} (s)}{\p s},
\ee
which allows to express $\mu_b$ in terms of $s$ and $t$ or $\tau$ and vice versa.
Thus, the disk amplitudes are related by the Legendre transform
\be
{\cal F}^o_{(0)} (s)={\cal F}^o_{(0)} (\mu_B)- s \mu_B.
\ee
The next orders of the $\bar g $-expansion can be obtained by computation of the Gaussian integral with perturbation, in particular
\be\label{from_mu_to_s_cylinder}
{\cal F}^o_{(1)} (s)={\cal F}^o_{(1)} (\mu_B)-\frac{1}{2}\log\left(\frac{\p^2 {\cal F}^o_{(0)} (\mu_B)}{\p \mu_B^2}\right),
\ee
where the expression for $\mu_{B}$ in \eqref{eq:2.35} is used.

Another, but essentially equivalent, 
version of the Laplace transform of GF of OIT was considered in \cite{BT_17}. It was shown
that after the Laplace transform the GF of OIT coincides with the Baker-Akhiezer function of the Kontsevich-Witten tau-function of the KdV hierarchy:
\be\label{wave_function}
e^{F^o(t,z)}=z^{-1/4}\,e^{\lambda^{-1}\sum_{k\geq 0} (t_k-\delta_{k,1}) \frac{z^{2k+1}}{(2k+1)!!}} e^{F^c\left(t_k-\lambda \frac{(2k-1)!!}{z^{2k+1}}\right)-F^c(t)},
\ee
where $z$ is to be identified with $i\sqrt{2\mu_B}$. 
In appendix B we compare the first term of this identity with our simplest result in section 4.1.
The same relation between the GF of open and closed versions of MG, which is probably the simplest example of the more fundamental relation between open and closed theories, was obtained in \cite{J_93}. This demonstrates,
that the Laplace transform \eqref{invLaplace} indeed provides a correct way to introduce the boundary cosmological constant into the OIT. Below,  we describe explicit computations, which also support this claim.

The conjectural relation through the Laplace transform allows us to translate the properties of the GF from $s$ to $\mu_B$ pictures and back. In particular, the Virasoro constraints 
\be
\label{Virasoro-constraint-mu_B}
{\cal C}_n~ e^{{\cal F}^c+{\cal F}^o(\mu_B)}=0\quad{\rm for}\quad n \ge -1,
\ee 
where \cite{MR_18}
\be
\label{open_Virasoro_generator_mu_B}
{\cal C}_n = {\cal L}_n + (-\mu_B)^n 
\left( -\mu_B\frac{\partial }{\partial \mu_B} 
- \frac{n+1}4  \right), %.
\ee
with $\cL_n$ being $L_n$, replacing $t_i$ with $\t_i$ except $t_1\to \t_1+1$,
can be obtained by the Laplace transform of Virasoro generators \eqref{open_Virasoro_generator} for the cases, where the GF depends on the infinite set of the KdV variables \cite{DW, DJMW_92,BT_17}.

Similarly, for  the open KdV hierarchy in $\mu_B$ space we have:
\be 
\label{BMG-KdV-mu_B}
\frac{2n+1}{2}\frac{\partial {\cal F}^o}{\partial \t_n}
 =- \mu_B \frac{\partial {\cal F}^o}{\partial \t_{n-1}}
 +\frac{\lambda^2}{2}\frac{\partial {\cal F}^o}{\partial \t_0}
 \frac{\partial^2 {\cal F}^c}{\partial \t_0\partial \t_{n-1}}
 -\frac{\lambda^2}{4}\frac{\partial^3 {\cal F}^c}{\partial \t^2_0\partial \t_{n-1}} 
 \quad{\rm for}\quad n\geq1.
\ee
The  open string equation becomes 
\be 
\label{MG-OSE-mu_B}
0=\sum_{n\geq0}\t_{n+1}\frac{\partial {\cal F}^o}{\partial \t_n} +\frac{\partial {\cal F}^o}{\partial  \mu_B}. 
\ee 
In addition, the  $s$-flow equation \eqref{OIT-s-flow}
has a new form in the $\mu_B$-space:  
\be
\label{MG-s-flow-mu_B}
-\mu_B 
= \lambda \left\{
\frac{1}{2}\left(\frac{\partial {\cal F}^o}{\partial \tau_0}\right)^2 
+ \frac12 \frac{\partial^2 {\cal F}^o}{\partial \tau_0^2}
+ \frac{\partial^2 {\cal F}^c}{\partial \tau_0^2}
\right \},
\ee 
which we call boundary condition equation (BCE).
The BCE provides a simple and consistent check of the GF 
in the presence of the boundary and 
the boundary cosmological constant plays a role of 
the boundary condition.

%%%%%%%%%%%%%%%%%%%%%%%%% 
\section{Generating function of minimal gravity with boundaries} \label{S3}
%%%%%%%%%%%%%%%%%%%%%%%%

%%%%%%%%%%%%%%%%%%%%%%%%% 
\subsection{Generating function on a disk}

According to one-matrix model approach, 
the continuum limit of the matrix variable 
is described by a differential operator   
$\hat{Q}_2= -\pd_x^2+u(x)$, where $u$ is given by \eqref{u} 
\cite{Doug_90,BDSS_90}, and the GF of BMG can be expressed in terms of $u$, obtained from GF of MG without boundary.
For the GF on a disk one may use the dispersionless limit  (neglecting 
derivatives) of   $\hat{Q}_2$, the second order polynomial in $y$; $Q_2= y^2+v(x)$.

The GF on a disk %is given symbolically by $\left \langle{\rm Tr} \log (\mu_B + Q_2) \right \rangle$ \cite{IR_11} and
has the integral representation \eqref{GF_disk}
with the proper normalization \cite{MSS_91}:
\bea
\label{eq:3.1}
{\cal F}^o_{(0)}(\tau, \mu_B)
&=
\frac{i}{\sqrt{2\pi^2}}\int_0^\infty \frac{dl}{l} \ e^{-l\mu_B} 
\int_{\tau_0}^\infty dx \int_{\mathbb{R}} dy \ e^{ -l ( y^2+v(x)) } 
\nonumber\\
&
=\frac {i}{\sqrt{2 \pi}}\int_0^\infty \frac{dl}{l^{3/2}} e^{-l\mu_B}
		\int_{\tau_{0}}^\infty dx \ e^{-l v(x)},
\eea 
and is given symbolically by $\left \langle{\rm Tr} \log (\mu_B + Q_2) \right \rangle$,
which is straightforwardly extendable to incorporate multiple boundaries 
and to impose boundary conditions \cite{IR_11}.
In \eqref{eq:3.1}, $v(x)$ is the solution of the string  polynomial equation 
\eqref{string_polynomial_2d_gravity}  with $\t_0$ replaced by $x$. 
As is discussed in the previous section, there are $p+1$ solutions to this equation, and to get the GF of BMG one have to take the solution 
$v(x)  \propto  \sqrt{- \tau_{p-1}}$
when all the parameters switched off except  $ \t_{p-1}$
($\propto \mu $).

This GF satisfies the lowest order  of the open KdV hierarchy 
in $\mu_B$-space \eqref{BMG_open_KdV_0} as we show in the appendix A.  
The open string equation \eqref{MG-OSE-mu_B} in the lowest order in $\lambda$
trivially follows from the string polynomial equation  
\eqref{string_polynomial_2d_gravity}. 
In addition, one can show that the lowest order of the BCE \eqref{MG-s-flow-mu_B},
\be 
\label{g=0-mu_B}
-\mu_B   = \frac12  \left(\frac{\pd {\cal  F}^o_{(0)}}{\pd \t_0} \right)^2
+\frac{\pd^2  {\cal  F}^c_{(0)} } {\pd \t_0^2},
\ee
 is also satisfied if one notes that
\be 
\label{g=0-mu_B2}
	\frac{\pd  {\cal  F}^o_{(0)}}{\pd \t_0}
	 =-i\sqrt{\frac{\mu_B+w}{2\pi}}\Gamma\left(-\frac12\right)=\sqrt{2}i\sqrt{\mu_B+w},
\ee
where $w=v(\t_0)$ denotes 
the relevant solution of ${\cal P} (\t, v) =0$. 

To simplify the expression one can change the integration variable $x$ into $v$  
if the string polynomial equation for ${\cal P}(\tau, v ) $ in \eqref{string_polynomial_2d_gravity} is used:  
\be 
\label{BMG-integral-string-polynomial}
{\cal F}^o_{(0)}(\tau, \mu_B)
	=-\frac {i}{\sqrt{2 \pi}}\int_0^\infty \frac{dl}{l^{3/2}} e^{-l\mu_B}
		\int_{w}^\infty dv \,  {\cal  P}^{(1)}(\t, v)  \,  e^{-l v}.
\ee
Here  $  {\cal P}^{(1)}(\tau,v) = d{\cal P} /dv $ 
plays the role of the Jacobian factor
$dx/dv = - d{\cal P} /dv $.
It should be emphasized that the integration variable $v$ is independent of $\t_n$.
After the integration by parts in $v$,
one gets
\be 
{\cal F}^o_{(0)}(\tau, \mu_B)
	=-\frac {i}{\sqrt{2 \pi}}\int_0^\infty \frac{dl}{l^{1/2}} e^{-l\mu_B}
		\int_{w}^\infty dv \,  {\cal  P}(\t, v)  \,  e^{-l v},
\ee
where we use ${\cal P} (\t, w) =0 $.
Therefore, one has 
\be
{\cal F}^o_{(0)}(\tau, \mu_B)
	=\sum_n \t_n \ll O_{n} \rr_{\rm disk},
\ee
where, due to ${\cal P} (\t, w) =0$, we have an identity
\be
\label{On}
	\ll O_{n} \rr_{\rm disk} =  \frac{\p {\cal F}^o_{(0)}}{\p \t_n}
	=	-\frac {i}{\sqrt{2 \pi}}\int_0^\infty \frac{dl}{l^{1/2}} e^{-l\mu_B}
		\int_{w}^\infty dv  \frac {v^n}{n!}  \ e^{-l v},
\ee 
and
\be
0= \sum_n \t_n  \frac{\p  \ll O_{n} \rr_{\rm disk}}{\p \t_m}.
\ee
The correlation functions $\ll O_{n} \rr_{\rm disk}$ depends only on $w$ and $\mu_B$.

To evaluate the integrals in \eqref{On}, 
we scale $v=w \eta$ 
and put the scale-free monomial $\eta^n$ 
as a linear combination of the Legendre polynomials
 $P_k$:
\be
	\eta^n=\sum_{k=n,n-2,\dots \geq0} (2k+1)\, n!  \, a_{n,k}\,  P_k(\eta),
\ee
where
 \be
	a_{n,k}=\frac{1}{2^{(n-k)/2}((n-k)/2)!(n+k+1)!!}.
\ee
Then the integration over $\eta$ is given as the  modified Bessel function of the second kind $K_{n}$:
\be
	  \int_1^\infty d\eta\,  {e^{-w l\eta}} P_k(\eta)
	  =\sqrt{\frac{2}{\pi w l }} \  {K_{1/2+k}(w l)} .
\ee 
Furthermore, the integration over $l$ is performed (after analytic continuation if necessary) to give 
\be
	\int_0^\infty \frac{dl}{l}\ e^{-l\mu_B} K_{1/2+k}(w l)=\frac{2\pi(-1)^{k+1}}{2k+1}\cosh((1/2+k)\, \theta).
\ee
Here we put  $\mu_B=w \cosh( \theta)$. 
As a result, \eqref{On} is given in terms of the Chebyshev polynomial $T_n (\cosh(x)) = \cosh(nx) $:
\be\label{disk:corr:numb}
	\ll O_{n} \rr_{\rm disk}
	=-i\,w^{n+1/2}(-1)^{n+1}\sum_{k=n,n-2,\dots\geq0}
a_{n,k} \ T_{2k+1}(\cosh(\,\theta/2)).
\ee 
It is noted that $\ll O_{n} \rr_{\rm disk}$ in general depends on the KdV parameters
since $w$ and $\theta$ are
functions of KdV parameters.
However, on-shell 
$w$ and $\theta$ reduce to certain constant values, 
$w \propto \sqrt{\mu}$ and $\cosh (\theta) \propto \mu_B /\sqrt{\mu}$.

%%%%%%%%%%%%%%%%%%%%%%%%% 
\subsection {Higher $\bar g$-expansion}\label{Hgexp}
%%%%%%%%%%%%%%%%%%%%%%

The open KdV hierarchy \eqref{BMG-KdV-mu_B}
and the open string equation \eqref{MG-OSE-mu_B}
allow one to further evaluate the higher  $\bar g \ge 1$  contributions using 
the $\bar{g}$-expansion:
\bea
\label{BMG-genus-expansion-KdV}
	\frac{2n+1}{2}\frac{\pd {\cal  F}^o_{(\bar g)}}{\pd \t_n}
	&=-\mu_B \frac{\pd {\cal  F}^o_{(\bar g)}}{\pd \t_{n-1}}
		+	
		\frac12 \sum_{\bar g_1+2 \bar g_2=\bar g}
		\left( 
		\frac{\pd {\cal  F}^o_{(\bar g_1)}}{\pd \t_0}
		\frac{\pd^2 {\cal  F}^c_{(\bar g_2)}}{\pd \t_0\pd \t_{n-1}}
		\right)
		-
		\frac14 \frac{\pd^3 {\cal  F}^c_{\left((\bar g-1)/2\right)}}{\pd \t^2_0\pd \t_{n-1}},	 
	\\
\label{BMG-genus-expansion-string}
	0&=\sum_{n\geq0}\t_{n+1}\frac{\pd {\cal  F}^o_{(\bar g)} }{\pd \t_n}+\frac{\pd {\cal  F}^o_{(\bar g)}}{\pd  \mu_B}.
 \eea
Here and below the term ${\cal  F}^c_{((\bar g-1)/2)}$ is present only when $\bar g$ is odd. 
 In addition  BCE  \eqref{MG-s-flow-mu_B} shows that  higher $\bar g\ge1$ equation becomes linear in ${\cal  F}^o_{(\bar g)}$:
\be
 \label{g>0-mu_B}
0=\sum_{\bar g_1 + \bar g_2=\bar g} \frac12  \frac{\pd {\cal  F}^o_{(\bar g_1)}}{\pd \t_0} 
\frac{\pd {\cal  F}^o_{(\bar g_2)}}{\pd \t_0} 
+\frac12 \frac{\pd^2  {\cal  F}^o_{(\bar g-1)}}{\pd \t_0^2}
+\frac{\pd^2  {\cal  F}^c_{\left(\bar g/2 \right)} } {\pd \t_0^2}.
\ee 
The $\bar g$-expansion shows that higher $\bar g$ contribution, more precisely, 
its first derivatives,  is given in terms of lower $\bar g$ solution.
Therefore, GF for each $\bar g$ can be obtained  from 
${\cal F}^o_{(0)} (\tau, \mu_B) $, 
which satisfies the lowest order non-linear equation  \eqref{g=0-mu_B}.   

One can obtain the GF for  $\bar g= 1$ if one uses  the
BCE \eqref{g>0-mu_B}
which simplifies for $\bar g= 1$ as the following: 
\be  
0= 
  \frac{\pd {\cal  F}^o_{(0)}}{\pd \t_0}   \frac{\pd {\cal  F}^o_{(1)}}{\pd\t_0}  
+\frac12 \frac{\pd^2  {\cal  F}^o_{(0)}}{\pd \t_0^2}
= \frac{\pd {\cal  F}^o_{(0)}}{\pd \t_0}   \frac{\pd}{\pd \t_0}  
\left( {\cal  F}^o_{(1)}+ \frac 12 \log  \left( \frac{\pd {\cal  F}^o_{(0)}}{\pd \t_0} \right) 
\right).
\ee 
The solution  has the form
\be 
\label{BMG-GT-g=1}
{\cal  F}^o_{(1)}=   - \frac 12 \log  \left( \frac{\pd {\cal  F}^o_{(0)}}{\pd \t_0} \right), 
\ee 
where  $\tau_0$-independent but 
$\t_{n>0}$-dependent contribution turns out to vanish
except the trivial constant \cite{MR_18}.

This solution can be compared with the GF of BMG from the matrix model computation on cylinder
$\F^o_{(1)} \propto\left \langle \Big( {\rm Tr} \log (M+ \mu_B) \Big)^{ 2} \right \rangle_c$,
where the subscript $c$ denotes the connected part
and $M$ is the hermitian matrix. 
In the continuum limit, 
 one has to replace $M$ by $Q_2 $  to have the form \cite{BDSS_90}: 
\bea
	\F^o_{(1)} 
	= \left(\frac{i}{\sqrt{2\pi^2}}\right)^2
	&  \int_{\tau_0}^\infty dx_1 \int_{-\infty}^{\tau_0} dx_2 	\int_0^\infty \frac{d l_1d l_2}{l_1l_2}
	\nonumber \\
	&\qquad \qquad  \times
	\ll x_1 \right| e^{-l_1 \left(\mu_B+w -\p^2 \right)}\left| x_2 \rr 
	 \ll x_2 \right| e^{-l_2 \left(\mu_B+ w -\p^2 \right)} \left| x_1 \rr,
\eea
where 
the integration range of $x_1$ does not overlap with that of  $x_2$ because of 
the connected part.  
In addition, $v(x)$ in $Q_2$  is replaced by $w$  because 
derivative of $v$ does not contribute to the cylindrical contribution \cite{MSS_91}. 
One may evaluate the integral easily, by inserting the identity 
$1=\int_{-\infty}^\infty dp_a\left| p_a \rr \ll p_a \right|$, with $\ll p_a | x_i\rr=e^{ip_ax_i}$,
and performing the $p_a$-integral (Gaussian integral), the matrix element is evaluated as
\be
	\ll x_i \right| e^{ l_i \, \p^2 } \left| x_j \rr
	=\sqrt{\frac{\pi}{l_i}} \, e^{-\frac{(x_i-x_j)^2}{l_i}}.
\ee
This shows that  \cite{MSS_91}
\be
	\F^o_{(1)} 
	=\frac{1}{4\pi}
	\int_0^\infty \frac{d l_1}{l_1^{1/2}} \int_0^\infty \frac{d l_2}{l_2^{1/2}} 
	\, e^{-(l_1+l_2) \left(\mu_B+w\right)}\frac{1}{l_1+l_2}.
\ee
To regularize this divergent integral we take a derivative with respect to $\t_n$:
\be
	\frac{\p \F^o_{(1)} }{\p \t_n}
	=-\frac{1}{4}\frac{\p w }{\p \t_n}
	\frac{1}{\mu_B+w}.
\ee
Integrating over $\t_n$ again gives 
\be
\label{GF-cylinder}
\F^o_{(1)} 
=   - \frac 14 \log  \left( \mu_B +w \right) +a,
\ee
where $a$ is the integration constant independent of $\t_n$.
If the constant $a$ is fixed as  $ -\frac12\log(\sqrt{2} i)$,
then the result is consistent with  
\eqref{BMG-GT-g=1} recalling \eqref{g=0-mu_B2}.
This demonstrates that the matrix calculation coincides with that of the open KdV results.
We expect this holds for higher $\bar g$ solution.

%%%%%%%%%%%%%%%%%%%%%%%%% 
\section{Generating function of intersection theory with boundaries} \label{S4}
%%%%%%%%%%%%%%%%%%%%%%%%%%

GF of the OIT satisfies the open KdV hierarchy \eqref{open_KdV} and the open Virasoro constraints \eqref{open_Virasoro_constraint}. So, one can evaluate this GF solving the open KdV hierarchy, or, equivalently, Virasoro constraints, using the initial conditions 
\be
\label{initial-condition-OIT}
F^o (t_0, t_{i>0}=0, s) = \frac1\lambda \left( st_0+\frac{s^3}{3!}  \right).
\ee
The solution is unique \cite{B_16}.

In this section, we will find the GF of OIT in a different way.
Namely, we demonstrate that the Laplace transform \eqref{Laplace} 
converts the GF of a particular topological solution of the BMG in $\mu_B$-space  into the GF of OIT in $s$-space:  
\be
\label{Laplace-OIT}
e^{{F}^o( s)} = 
\int d \mu_B \, e^{-s\mu_B/\lambda } ~ e^{{F}^o(\mu_B)}.
\ee
If the GF obeys the open KdV hierarchy 
and the GF on a disk ($\bar g= 0$ contribution) 
satisfies the initial condition \eqref{initial-condition-OIT},
then, higher $\bar g>0$ contribution should also work 
due to the uniqueness of the solution.  

%%%%%%%%%%%%%%%%%%%%%%%%% 
\subsection{Generating function on a disk} 

Note that the Laplace transform  \eqref{Laplace-OIT} 
reduces to the Legendre transform for $\bar g=0$ \cite{MR_18}:
\be
\label{Legendre}
 F^o_{(0)} (t, s) 
= F^o_{(0)}(t, \mu_B) -  s\mu_B,
\ee
and the boundary parameters, $s$ and $\mu_B$, are related through 
\be 
\label{s-muB}
s =\frac{\partial F^o_{(0)}(t, \mu_B)}{\partial \mu_B} \quad
{\rm or}\quad
\mu_B =-\frac{\partial F^o_{(0)}(t ,s) }{\partial s}.
\ee 
To get the GF of OIT we have to take a particular GF on the $\mu_B$ side, $F^o_{(0)}(t, \mu_B)$. We claim that it is given by the disk GF \eqref{GF_disk} where $\tau_0$ substituted by $t_0$, 
and $v$ is the solution of the polynomial string equation \eqref{IT-string-polynomial} as $p$ tends to infinity. 
So GF of  OIT  is represented by 
\eqref{BMG-integral-string-polynomial}
but with replacing the string polynomial 
 ${\cal P} (\t, v) $ with $ P (t, v) $ in \eqref{IT-string-polynomial}:
\be 
\label{GF_0-OIT-mu_B}
F^o_{(0)}(t, \mu_B)
	=-\frac {i}{\sqrt{2 \pi}}\int_0^\infty \frac{dl}{l^{3/2}} e^{-l\mu_B}
		\int_{w}^\infty dv \,  P^{(1)}( t, v)  \ e^{-l v}.
\ee
Here $w$ corresponds to the solution of
$P (t, v) =0 $, regarding 
the terms with the parameter set $\{t_k\}$ is treated as a perturbation.

Let us check if this  OIT solution
coincides with the GF in $s$-space. 
We use the Legendre transform
with the conjugate variable $s$  
\be
\label{OIT-s-mu_B-g=0}
s   =\frac{\partial F^o_{(0)} }{\partial \mu_B}
 = \frac {i}{\sqrt{2 \pi}}\int_0^\infty \frac{dl}{l^{1/2}} e^{-l\mu_B}
		\int_{w}^\infty dv \, P^{(1)}(t, v)  \ e^{-l v}.
\ee 
To get some idea about how to find the explicit form of $F^o_{(0) } (t, s )$, we start with $p=1$ case and move on to $p=2,3$ and $p=4$,
and then extract generic features. Of course, for any $p$ the case with $p-1$ can be obtained if one puts $t_{p-1}=0$.

%%%%%%%%%%%%
\paragraph{$p=1$ case:} In this case only $t_0$ is present, $P^{(1)}=-1$ and $w=t_0$. 
Thus \eqref{GF_0-OIT-mu_B}  has the simple form:
\be
F^o_{(0)} (t_0, \mu_B)
=\frac {i}{\sqrt{2\pi}}\int_0^\infty \frac{dl}{l^{5/2}}
 e^{-l\left(\mu_B+  w\right)}.
\ee
Note that this integral depends only on the sum $\mu_B+  w$.
This integral is divergent as $l \to 0$ and needs regularization to be finite. 
We note the differentiation which makes the integral finite:
\be
	\frac{\p^2 F^o_{(0)}}{\p \mu_B^2 }
	=\frac {i}{\sqrt{2\pi}}\int_0^\infty \frac{dl}{l^{1/2}} e^{-l\left(\mu_B+  t_0 \right)}
	=\frac {i}{\sqrt{2}} (\mu_B+  w)^{ -1/2}.
\ee
 After integration over $\mu_B$ once,
we have
\be
\label{OIT-s-g=0}
	s= \frac{\p F^o_{(0)}}{\p \mu_B }
	=i\sqrt{2}  (\mu_B+  w)^{1/2},
\ee
discarding $\mu_B$ independent term.  
Likewise, one more integration gives 
\be
\label{Fo-muB-space}
F^o_{(0)} (t_0, \mu_B) =i\frac{2\sqrt{2}}{3}  
(\mu_B+ w)^{3/2}.
\ee
We provide another derivation from \eqref{wave_function} in appendix B.

After the Legendre transformation, we have 
GF of OIT:
\be
\label{Fo-s-space}
F^o_{(0)} (t_0,s)=s w+ \frac{s^3}{3!} = st_0 + \frac{s^3}{3!} ,
\ee
which is exactly the same as the initial condition \eqref{initial-condition-OIT}. 

%%%%%%%%%%%%
\paragraph{$p=2$ case:}

In the presence of  two variables $t_0$ and $ t_1$, one has 
$ P^{(1)}(t,v) = t_1-1$,  which we also denote by $-\xi_1 $, and    $w=t_0/(1-t_1)$.
As in the $p=1$ case, we can evaluate
\eqref{GF_0-OIT-mu_B} after regularization:
\bea
\label{eq:4.12}
s=\frac{\p F^o_{(0)}}{\p \mu_B }
&=i\sqrt{2}\,\xi_1  (\mu_B+  w)^{1/2},
\\
F^o_{(0)}(t, \mu_B)
&=i\frac{2\sqrt{2}}{3}\,\xi_1 \left(\mu_B+ w\right)^{3/2}.
\eea
The Legendre transform results in GF 
\be
\label{eq:4.14:F_0}
F^o_{(0)}(t_0,t_1, s)
= \frac{st_0}{1-t_1}+ \frac{s^3}{3!(1-t_1)^2}
= sw	+ \frac{s^3}{3!(1-t_1)^2},
\ee
which is in the agreement with the results of \cite{PST_14}.

As the number of KdV  parameters increases, the evaluation becomes not easy to carry out. 
To simplify it, 
we note that the open string equation \eqref{OIT-string} 
at $\bar g=0$ has the form of differential equation:
\be\label{OIT-string-g=0}
D F^o_{(0)}=s,
\ee
where
\be
\label{OIT-string-g=0-op}
D \equiv 
\frac{\partial }{\partial t_0} 
- \sum_{n \ge 0} t_{n+1} \frac{\partial }{\partial t_n}.
\ee 
The inhomogeneous solution 
to the equation \eqref{OIT-string-g=0}
is $sw$.  This can be seen as follows.
Since $w$ is the solution of the string polynomial equation $P(t, v)=0$, 
its derivative $\p P(t, w(t))/ \p t_0 =0$ satisfies an identity:
\be
\frac{\pd w }{\pd t_0}=\sum_{n\geq0} 
t_{n+1}\frac{\pd  w }{\pd t_n}+1.
\ee 
Thus $sw$ satisfies \eqref{OIT-string-g=0}. 
In addition, there exist solutions of the corresponding
homogeneous equation. 
A homogeneous solution $f$ can be  
put as a function of   $ s$ and  
a convenient set of parameters  
$\xi_1, \cdots, \xi_{p-1}$:
\be
\xi_n =- \frac{d ^n P(t, v)}{dv^n}\bigg|_{v=w} \equiv -P^{(n)}(w)~ ~~{\rm for}~~n=1, \cdots p-1.
\ee 
One may easily show that $D\, \xi_n =0$.  
Therefore, we may put GF at $\bar g=0$ as the following form:
\be 
\label{structure-OIT-GF-g=0}
 F^o_{(0) } (t, s ) = s w + f(s, \xi),
\ee 
with the homogeneous solution $f$. This structure of GF is  already seen
 in $p=1$ and 2 cases. 
 
 The variables $\xi_n$ are well known in the matrix models \cite{AJM,ACKM}. These variable are extremely convenient for investigation of the GF and correlation functions for the resolvents for the closed case. As the GF of the BMG can be related to the correlation function of the local operators in MG \cite{BDSS_90,MSS_91},  it is not very surprising that they also show up in the theory with boundary.

%%%%%%%%%%%%
\paragraph{$p=3$ case:}

For  the case $p=3$, the solution of the polynomial string equation $w$  has the form
\be
	w
	=\frac{1-t_1}{t_2}\left(1-\sqrt{1-\frac{2t_0t_2}{(1-t_1)^2}}\right)
	=\frac{t_0}{1-t_1}\left(\frac{2}{1+\sqrt{1-\frac{2t_0t_2}{(1-t_1)^2}}}\right).
\ee  
One finds
\be
\label{s-mu_B-p=3}
s =\frac{\p F^o_{(0)}(t,\mu_B)}{\p  \mu_B} 
=\sqrt{2}\,i\,(\mu_B+w)^{1/2} \left( \xi_1
-\frac{2}{3}\xi_2(\mu_B+w)\right),
\ee
where (assuming $ t_1<1$)
\bea \label{xiparam}
\xi_1 &=-P^{(1)} (t, w) =1- t_1- t_2 w 
=\sqrt{ (1-t_1)^2 - 2 t_0 t_2 },\\
\xi_2&=-P^{(2)} (t, w)=-t_2.
\eea
Integrating over $\mu_B$, one has 
\be
  F^o_{(0)}(t,\mu_B) 
=\sqrt{2}\,i\,(\mu_B+w)^{3/2} \left( \frac 23 \xi_1 - \frac{4}{5!!}\xi_2(\mu_B+w)\right).
\ee
Noting that   ${\p \xi_1}/{\p t_0}  =  {\xi_2}/{\xi_1}$,
one can check that $ F^o_{(0)} $ satisfies the BCE equation \eqref{g=0-mu_B}.

To find GF in $s$-space, we put $\mu_B$ in powers of $s$ 
by solving \eqref{s-mu_B-p=3}:
\be
\label{h_0-p=3}
	\mu_B =-w-\frac{s^2}{2\xi_1^2}\sum_{n=0}^\infty a_n z_1^n;
	\qquad
	a_n=\frac{(-1)^n}{n+1}\binom{3n+1}{n},
\ee 
with  $z_1= {s^2 \xi_2}/{(3\xi_1^3)}$.
Then, using the second relation in \eqref{s-muB},
one can easily find GF directly by integrating over $s$:
\be
\label{GF-p2}
	F^o_{(0)}(t, s)
	=sw 	+\frac{s^3}{2\xi_1^2}
	\sum_{n=0}^\infty\frac{a_n z_1^n}{(2n+3)},
\ee
which has the expected structure 
of \eqref{structure-OIT-GF-g=0}. 

%%%%%%%%%%%%
\paragraph{$p=4$ case:}

For the case $p=4$ we have  
\be
\label{OIT-s-p=4}
s =\frac{\p F^o_{(0)}}{\p  \mu_B}=  \sqrt{2}\,i\,\left(\mu_B+w\right)^{1/2}\,
	 \left( \xi_1- \frac2{3!!}\xi_2(\mu_B+w)+ \frac{4}{5!!}\xi_3({\mu_B}+w)^2\right).
\ee
We note that \eqref{OIT-s-p=4} is a polynomial equation of the form
\be
\label{polynomial-h_0-p=4}
	\frac1 {h_0} = (1+ z_1 \, h_0+ z_2 \, h_0^2 )^2,
\ee
where $h_0=- {2\xi_1^2(\mu_B+w)}/ {s^2}$, $z_1= {s^2\xi_2}/{(3\xi_1^3)}$ and $z_2= {s^4\xi_3}/{(15\xi_1^5)}$.
One can find $h_0$ as a series in $z_i$'s: 
\be
\label{h_0-p=4}
	h_0
	=\sum_{n,m=0}^\infty a_{n,m} \, z_1^n z_2^m; 
	\qquad 
 	a_{n,m}
	=2\frac{(-1)^{n+m}(3n+5m+1)! }{n!\,m!\,(2n+4m+2)!},
\ee
where $a_{n,0}=a_n$ in \eqref{h_0-p=3},
so that   the solution  \eqref{h_0-p=4}
reduces to the one in \eqref{h_0-p=3}
when $z_2 \to 0$.
This shows that
\be
\mu_B
	= -w -\frac{s^2}{2 \xi_1^2} 
	\sum_{n,m=0}^\infty a_{n,m} \, z_1^n z_2^m.
\ee
Using  the  second relation in \eqref{s-muB}
one integrates $\mu_B$ over $s$ to find  GF of the form
\be
	F^o_{(0)}(t, s)
	=sw 	+\frac{s^3}{2\xi_1^2}
	\sum_{n,m=0}^\infty\frac{a_{n,m} \, z_1^n z_2^m}{(2n+4m+3)}.
\ee 

%%%%%%%%%%%%
\paragraph{Arbitrary $p$:}

In general, \eqref{OIT-s-mu_B-g=0} provides the relation between 
$\mu_B$ and $s$ as follows:
\be
\label{eq:relation:s-muB:explicit}
s=
\frac{\p F^o_{(0)}}{\p  \mu_B}= \sqrt{2}\,i\, \sum_{n=0}^{p-2}\frac{(-2)^n\xi_{n+1} (\mu_B+w)^{n+1/2}}{(2n+1)!!} .
\ee
We find $F^o_{(0)}(t, \mu_B)$  
by integrating \eqref{eq:relation:s-muB:explicit} over $\mu_B$.
\be
\label{OIT-GF-g=0:muB}
F^o_{(0)}(t, \mu_B) 
= -\sqrt{2}\,i\,\sum_{n=0}^{p-2} 
\frac{(-2)^{n+1}\xi_{n+1} (\mu_B+w)^{n+3/2}}{(2n+3)!!}.
\ee
Similar to \eqref{polynomial-h_0-p=4}, we 
rewrite \eqref{eq:relation:s-muB:explicit} in the following form
to find  $\mu_B$ in power series of $s$
\be
\label{eq:relation:s-muB}
\frac1 {h_0} = \left( 1+ \sum_{n=1}^{p-2}z_n \,  h_0^n  \right)^2,
\qquad    z_i=\frac{s^{2i}\xi_{i+1}}{(2i+1)!!\xi_1^{2i+1}},
\ee
where 
$  h_0$ is the same in  \eqref{polynomial-h_0-p=4}: $h_0=- {2\xi_1^2(\mu_B+w)}/ {s^2}$.
Then $h_0$  is given in a power series of $z_k$'s: 
\be 
\label{solution-h}
h_0 =\sum_{n_k\ge 0} a_{n_1, \cdots, n_{p-2}} 
\  z_1^{n_1}  \cdots z_{p-2}^{n_{p-2}},
\ee
where the coefficient $a_{n_1, \cdots, n_{p-2}}$ 
has the form
\be 
a_{n_1,n_2,\dots,n_{p-2}}
	=2\frac{(-1)^{n_1+n_2+\dots+n_{p-2}}(1+3n_1+5n_2+\dots+(2p-3)n_{p-2})! }{n_1!\,n_2!\,\dots\,n_{p-2}!\,
	(2+2n_1+4n_2+\dots+2(p-2)\,n_{p-2})!}.
\ee
By noting 
\be\label{mu_in_s}
\mu_B
	= -w -\frac{s^2}{2 \xi_1^2} \sum_{n_k\ge 0} a_{n_1, \cdots, n_{p-2}} 
\  z_1^{n_1}  \cdots z_{p-2}^{n_{p-2}},
\ee
one has GF in power series of $z_k$'s
by integrating over $s$ following the second relation in \eqref{s-muB}:
\be
\label{OIT-GF-g=0}
	F^o_{(0)}(t, s)
	=sw 	+\frac{s^3}{2\xi_1^2}
	\sum_{n_i=0}^\infty\frac{a_{n_1, \cdots, n_{p-2}} \, z_1^{n_1}  \cdots z_{p-2}^{n_{p-2}} }
	{(3+2n_1+4n_2+\cdots + 2(p-2)\,n_{p-2})}.
\ee 

To get the complete GF of OIT one should tend $p$ to infinity.
This GF \eqref{OIT-GF-g=0} 
is consistent with the theorem provided by Pandharipande et.\ al.\ \cite{PST_14}
which uses the limit $t_0\to0$. In this limit, one has 
$w\to0, \ \xi_1\to 1-t_1,\ \xi_i\to-t_i \ (i\geq2)$ and 
GF has the following form:
\be
\begin{aligned}
 	F^o_{(0)}(t, s)\big|_{t_0=0}
	&=
	\sum_{n_i=0}^\infty
	\frac{(1+3n_1+5n_2+\dots)!}{n_1!\,n_2!\,\dots\,
	(3+2n_1+4n_2+\dots)!}
	\times
	\\&\qquad\quad
	\times
	\left(\frac{t_{2}}{3!!}\right)^{n_1} 
	\left(\frac{t_{3}}{5!!}\right)^{n_2} 
	 \cdots 
	 \frac{s^{3+2n_1+4n_2+\dots}}
	 {(1-t_1)^{2+3n_1+5n_2+\dots}}.
\end{aligned}
\label{GF-PST} 
\ee
This provides the correlation numbers 
\be
	\langle O_{\alpha_1} \dots O_{\alpha_\ell} \sigma^k \rangle^o_0
	=\frac{\p^{\ell+k}F^o_{(0)}}{\p t_{\alpha_1}\dots \p t_{\alpha_\ell}\p s^k}\bigg|_{(t,s)=0}
	=\frac{(1 +\sum_{i=1}^\ell( 2 \alpha_i-1))!}{\prod_{i=1}^\ell (2\alpha_i-1)!!}.
\ee
Here we note that  the number $k$ of marked points on the boundary
is specified by the set $\{\alpha_i\}$: 
$k=3+\sum_{i=1}^\ell2(\alpha_i-1)$, 
which is clearly seen in \eqref{GF-PST}.

%%%%%%%%%%%%%%%%%%%%%%%%% 
\subsection {Higher $\bar g$-expansion}
%%%%%%%%%%%%%%%%%%%%%%

Universal formula for the GF on the cylinder ($\bar g =1$) in the $\mu_B$ picture
$F^o_{(1)}(t, \mu_B)$, was obtained in section \ref{Hgexp},  \eqref{GF-cylinder}:  
\be
\label{GF-OIT-cylinder}
F^o_{(1)} (\mu_B)
=   - \frac 14 \log  \left( \mu_B +w \right) + {\rm constant}.
\ee
We expect that this relation as well as the disk GF \eqref{OIT-GF-g=0:muB} and higher $\bar g$ contributions can be also extracted from the $\lambda$-expansion of the wave function formula \eqref{wave_function}. 

For the cylinder the relation between the GF in $s$ and $\mu_B$ pictures is given by
\eqref{from_mu_to_s_cylinder}: 
\be 
\label{OIT-GF-g=1}
	{F}^o_{(1)}( s)
	= 	   
	F^o_{(1)}( \mu_{B} )
	-\frac 12\log\left( \frac{\d^2 F^o_{(0)}( \mu_{B})} {\d  \mu_{B}^2}\right),
\ee
which, with the help of \eqref{OIT-GF-g=0:muB}, reduces to 
\be
\label{cyl_GF_exact}
{F}^o_{(1)}(s)=\left. -\frac{1}{2}\log\left(\xi_1 + \sum_{n=1}^{\infty}\frac{(-2)^n\xi_{n+1} (\mu_B+w)^{n}}{(2n-1)!!}\right)\right|_{\mu_B=\mu_B(s)} + c,
\ee
where $c$ is a constant.
This expression has the power series expansion in $z_i$'s 
if one uses the expression for $\mu_{B}$ in \eqref{mu_in_s}.

To find the constant $c$ we consider the case with $t_k=0$ for $k>2$. 
The disk amplitude for this case is given by \eqref{eq:4.14:F_0}, so the cylinder GF \eqref{cyl_GF_exact} is
\be
\label{GF-OIT-p=2_frg}
{F}^o_{(1)}(t,s)= -\frac{1}{2}\log\left(\xi_1\right) + c= -\frac{1}{2}\log\left(1-t_1\right) + c.
\ee
From this expression we can conclude, in particular, that $c=0$.

For $p=3$ ${F}^o_{(1)}( s)$ gives, up to logarithmic term, the  power series expansion in $z_1$ 
\bea
{F}^o_{(1)}(s)
&= -\frac{1}{2}\log\left(\xi_1 -2\xi_{2} (\mu_B+w)\right)\\
&=-\frac12  \log (\xi_1)
	-{{3}\over{2}}z_1+{{21}\over{4}}z_1^2-24z_1^3+{{981}\over{8}}z_1^4-{{6663}\over{10}}z_1^5
	+3765z_1^6+\cdots.
\label{GF-OIT-p=3_frg}
\eea

Let us compare this result 
with a solution to the open string equation and  the open KdV in $\bar g$-expansion in the $s$ picture.
The open string equation with  $ \bar{g} \ge 1 $ 
becomes a homogeneous differential equation
\be
D F^o_{(\bar g)} =0 ,
\ee
where $D$ is defined by \eqref{OIT-string-g=0-op}.
Therefore, GF with   $\bar g \ge 1 $  
is represented in terms of  $\xi_i$, the solutions of the homogeneous equation:
 \be
 \label{open-string-solution-higher-g}
 F^o_{(\bar g)}=F^o_{(\bar g)}(\xi_i, s) \quad{\rm  for}\quad \bar g\geq1.
 \ee 
 In addition, the  $s$-flow equation \eqref{OIT-s-flow}
in $\bar g$-expansion has the form
\be 
\label{eq:s-flow:}
\frac{\partial F^o_{(\bar g)}}{\partial s} 
=  \sum_{\bar g_1+\bar g_2=\bar g}   \frac12 
\left(  
 \frac{\pd F^o_{(\bar g_1)}}{\pd t_0} 
 \frac{\pd F^o_{(\bar g_2)}}{\pd t_0}
\right)
+ \frac12 \frac{\partial^2 F^o_{(\bar g-1)}}{\partial t_0^2}
+ \frac{\partial^2 F^c_{(\bar g/2)}}{\partial t_0^2},
\ee
which can be used to restrict further the form 
of \eqref{open-string-solution-higher-g}. 
Finally, it 
should satisfies open KdV equation: 
\bea 
\frac{2n+1}{2}\frac{\pd F^o_{(\bar g)}}{\pd t_n}
&=\sum_{\bar g_1+\bar g_2=\bar g}  
\left(  
 \frac{\pd F^o_{(\bar g_1)}}{\pd s}  
\frac{\pd F^o_{(\bar g_2)}}{\pd t_{n-1}}
\right)
+ \frac{\pd^2 F^o_{(\bar g-1)}}{\pd s \d t_{n-1}} 
\nonumber\\
&\qquad
+	\frac12 \sum_{\bar g_1+2 g_2=\bar g}
		\left( 
		\frac{\pd  F^o_{(\bar g_1)}}{\pd t_0}
		\frac{\pd^2 F^c_{( g_2)}}{\pd t_0\pd  t_{n-1}}
		\right)
		-
		\frac14 \frac{\pd^3 F^c_{\left((\bar g-1)/2\right)}}{\pd t^2_0 \pd t_{n-1}}
\quad{\rm for}\quad n \ge 1.
\eea

It is to be noted that 
$F^o_{(\bar g)}$ has an important parity property in $s$
\be
\label{OIT-parity}
F^o_{(\bar g)}(\xi_i,-s) = (-1)^{\bar g+1} F^o_{(\bar g)}(\xi_i, s).
\ee
The parity property is already seen  in \eqref{OIT-GF-g=0} of section 4.2
where GF $F^o_{(0)}(t, s)$   is odd in $s$.
The general proof can be done using the dimension 
of the moduli space given in \eqref{OIT-def}:
$\bar{g}+k$ must always be odd \cite{ABT}. 
Since $k$ denotes the power of $s$
in GF, one concludes  \eqref{OIT-GF-g=0}. 
The parity property is also consistent with the 
SD  considered in IT. 
Since $F^o$ is  scale-free
as seen in \eqref{open_Virasoro_constraint},
each term in $\bar g$-expansion is also scale-free. 
Noting that SD of  $\lambda$ is 3/2, 
one has SD of  $F^o_{(0)}$ is 3/2,
which is obvious in \eqref{OIT-GF-g=0}.
(SD of $s$ and $w$ are  1/2 and 1, respectively
and SD of $\xi_i$ and  $z_i$  are 0).
Therefore, the $\bar g$-expansion of $F^o$  shows that 
SD of  $F^o_{(\bar g)}$  is  $3(\bar g+1)/2$.
This shows that SD of  GF with $ \bar g$ even 
is a half-odd integer.
Since SD of  KdV parameter $t_n$ is an integer, 
the only way to have the half odd integer SD is
the quantity  proportional to odd power of $s$
which is reflected in the parity \eqref{OIT-parity}.

Let us provide a few simple checks of the GF expression \eqref{cyl_GF_exact}.
%%%%%%%%%%%%
\paragraph{$p=2$ case:}

At  $\bar g=1$, the open string equation shows that 
${\pd F^o_{(1)}}/{\pd t_0}=0$. 
Applying this condition to the $s$-flow equation, with the help of \eqref{eq:4.14:F_0} one has 
$\ {\pd F^o_{(1)}}/{\pd s} =0$.
This result is consistent with that $ F^o_{(1)}$ is the even function of $s$.  
Based on this fact, the open KdV has the form
\be 
	\frac{3}{2}\frac{\pd F^o_{(1)}}{\pd t_1}
=\frac{\pd^2 F^o_{(0)}}{\pd s \d t_{0}}
	 -\frac{1}{4}\frac{\pd^3 F^c_{(0)}}{\pd t_0^{\ 3}} 
=\left(1-\frac{1}{4}\right)\frac{\pd w}{\pd t_0 }
=\frac34 \frac1{(1-t_1)}.
\ee
The solution is given as 
\be
\label{GF-OIT-p=2}
	F^o_{(1)}=-\frac12\log(1-t_1) ,
\ee 
which coincides with \eqref{GF-OIT-p=2_frg}.

%%%%%%%%%%%%
\paragraph{$p=3$ case:}
%%%%%%%%%%%%

In this case, noting that  $F^o_{(1)}$  is scale-free
and an even function of $s$, we can find it
in terms of scale-free parameters, $\xi_1$ 
and $z_1$. A direct evaluation shows that
$F^o_{(1)}$ given by
\be
\label{GF-OIT-p=3}
	F^o_{(1)}
	=-\frac12  \log (\xi_1)
	+ \sum_{n=1}^\infty \sum_{k=0}^{n-1}
	\frac{3(-1)^n}{2n}\binom{3k}{k}\binom{3n-3k-2}{n-k-1}  z_1^n,
\ee
solves the $s$-flow equation, coincides with \eqref{GF-OIT-p=3_frg},
 and its expansion in a power series in $t_i$ and $s$ gives
\be
	F^o_{(1)}=
	{{t_{1}}\over{2}}+{{t_{1}^2}\over{4}}+{{s^2t_{2}}\over{2}}+{{t_{0}t_{2}}\over{2}}
	+{{3s^2t_{1}t_{2}}\over{2}}+t_{0}t_{1}t_{2}+3s^2t_{1}^2t_{2}+{{3t_{0}t_{1}^2t_{2}}\over{2}}
	+\dots,
\ee
reproducing the result provided explicitly by \cite{A2} under an appropriate identification of the parameters.

%%%%%%%%%%%%%%%%%%%%%%%%%%%%%%%%%% 
\section{Summary and discussion} \label{S5}

We investigated the relation between 
the two-dimensional minimal gravity 
(Lee-Yang series) on Riemann surfaces
with boundaries ($\mu_B$, 
the boundary cosmological constant)
and open intersection theory
($s$, the source of the boundary marked point). 
The generating functions of both theories are considered as 
solutions to the open KdV hierarchy and string equation. 
Since there are many solutions 
to the open KdV hierarchy with different analytic properties, 
one needs a proper way to identify the right solution.

We use the conjecture that the generating function
of the minimal gravity with $\mu_B$
and that of the intersection theory with $s$
is related by the Laplace transform.
The generating function on a disk
corresponds to the leading contribution
to the Laplace transform,
which reduces to the Legendre transform.
We obtain the generating function of the intersection 
theory from that of the minimal gravity using  the Legendre transform 
and confirm that the generating function of each theory 
belong to a different solution sector 
of the open KdV hierarchy and string equation.    
Based on this, we provide a systematic way 
to find the generating function in  $\bar g$-expansion,
$\bar g=0$ representing the disk.
($\bar g$ denotes the genus of the doubled Riemann surfaces,
equivalent to the Euler characteristic expansion).
As a non-trivial example of the Laplace transform, 
we further provide an explicit generating function of 
open intersection theory on a cylinder ($\bar g=1$),
from that of the minimal gravity
through the Laplace transform.

Higher $\bar g$-expansion is a more challenging problem.  
It will be interesting to find the generating function of the intersection theory 
through the Laplace transform
and compare it with  the combinatoric expression in $s$-space
for the all-genera generating function \cite{T_15}.
It is to be noted that a given term of the $\bar g$-expansion
contains contributions from the several topologically distinct surfaces. 
For example, $\bar g=2$ contains 
two different geometries: pants and kettle. 
For the open intersection theory,
the contributions of different types of surfaces
can be traced by the extension of the generating function \cite{A3,ABT}. 
However, the computations of the generating function of the minimal gravity with boundaries with topological structure different form the sphere with arbitrary number of boundaries is still not known,
but can be extracted from the relation in terms of the closed GF \cite{DJMW_92,J_93} or the matrix model computations \cite{MSS_91,BDSS_90}.

It is clear that the correlation functions in $\mu_B$-space 
presents the non-analytic behavior (square root branch),
which is useful to describe the correlation numbers of primary operators. 
On the other hand, the correlation functions of the intersection 
theory in $s$-space shows the polynomial behavior 
and are suited to describe the correlation numbers of the descendants. 
It is interesting to note that the very different role of 
the two theory spaces when related with the Laplace transform   
was used  in \cite{Kawai_13}
to solve  the cosmological problem
using the cosmological constant and its conjugate variable.

Another interesting  problem is the description of  the boundary 
gravitational descendants in open intersection theory introduced in \cite{B_15,B_16} 
and further investigated in \cite{A1,A2,A3,ABT}. 
This extended theory has a nice Kontsevich-Penner matrix model description. 
This identification immediately leads to the integrability of the 
deformed model, which is shown to be the tau-function of the KP hierarchy. 
It would be interesting to find the minimal gravity counterpart of this  deformed model. 
From the point of minimal gravity, 
this should correspond to consideration of 
the gravitating operators on boundaries and different types of the boundary conditions.  
The meaning of the Laplace transform from the 
point of view of the matrix integrals is not clear at the moment.

Finally, one may expect that the idea of this paper 
can be extended to the case of the so-called r-spin open intersection numbers \cite{BCT_18}.
One may relate this theory with 
$M(q, p)$ series of minimal gravity 
in terms of the $A_{q-1}$ Frobenius manifolds 
\cite{BR_15, ABR_17, BMR_DUAL_18,BY_15,AB_18}. 
It would be interesting to apply the Laplace transform to investigation of the open $p-q$ duality.
We are going to come back to these topics in the future publications. 

%%%%%%%%%%%%%%%%%%%%%%%% 
\subsection*{Acknowledgements}
The work of Muraki and Rim was partially supported by 
National Research Foundation of Korea  grant number 2017R1A2A2A05001164,
and of Alexandrov  by IBS-R003-D1 and by RFBR grant 17-01-00585. A.A. would also like to thank Vasily Pestun for his hospitality at IHES (supported by the European Research Council under the European Union's Horizon 2020
research and innovation programme, QUASIFT grant agreement 677368), where this project was completed.

%%%%%%%%%%%%%%%%%%%%%%%%
\appendix

\section{Open KdV of minimal gravity on a disk}
%%%%%%%%%%%%%%%%%%%%%%%%

We provide a simple way to prove that 
GF \eqref{GF_disk} satisfies  the openKdV hierarchy \eqref{BMG_open_KdV_0}.
According to \eqref{GF_disk}, 
\be  
\frac{\p {\cal F}^o_{(0)}}{\p \t_{n-1}}  
	=-\frac {i}{\sqrt{2 \pi}}\int_0^\infty \frac{dl}{l^{1/2}} e^{-l\mu_B}
		\int_{\tau_{0}}^\infty dx \,\frac{\p v}{\p \t_{n-1}}  \ e^{-l v(x)} ,
\ee 
where we put $\frac{\p v}{\p \t_{0}}=\frac{\p v}{\p x}$,
and using the string equation \eqref{MG-hierachy-g=0} we have  
\be 
\label{MG-one-point-g=0}
  \frac{\partial{\cal F}^o_{(0)}}{\partial \t_{n}}
=-\frac {i}{\sqrt{2 \pi}}\int_0^\infty \frac{dl}{l^{1/2}} e^{-l\mu_B} 
 \int_{w}^\infty dv\,  \frac {v^{n}}{n!}  \ e^{-l v},
\ee		
where we change the integration variable from $x$ to $v$.
If one multiplies  \eqref{MG-one-point-g=0} by $\mu_B$
(and  uses $\t_{n-1} $ instead of $\t_n$ for later convenience), 
the result can be put in terms of the derivatives of $l$: 
\be 
 -\mu_B \frac{\partial{\cal F}^o_{(0)}}{\partial \t_{n-1}}
=	-\frac {i}{\sqrt{2 \pi}}\int_0^\infty \frac{dl}{l^{1/2}} \left(\frac{\pd e^{-l\mu_B}}{\pd l}\right)
		\int_{w}^\infty dv  \frac {v^{n-1}}{(n-1)!}  \ e^{-l v}.
\ee
Using the integration by parts of $l$ one has 
\be 
 -\mu_B \frac{\partial{\cal F}^o_{(0)}}{\partial \t_{n-1}}
  =	-\frac {i}{\sqrt{2 \pi}}\int_0^\infty \frac{dl}{l^{1/2}}e^{-l\mu_B} \left(v+\frac1{2l}\right)
		\int_{w}^\infty dv  \frac {v^{n-1}}{(n-1)!}  \ e^{-l v},
\ee
where the surface term vanishes. 
We may subtract $ \frac{\partial{\cal F}^o_{(0)}}{\partial \t_{n}}$ from the above: 
\bea
&
 -\mu_B \frac{\partial{\cal F}^o_{(0)}}{\partial \t_{n-1}} 
 -\left(n+\frac12\right)  \frac{\partial{\cal F}^o_{(0)}}{\partial \t_{n}}
\nonumber\\
&\qquad\qquad=- \frac12\frac {i}{\sqrt{2 \pi}}\int_0^\infty \frac{dl}{l^{1/2}}
 e^{-l\mu_B} \frac{1}{l} \left(1 - \frac{v l}{n}\right)
		\int_{w}^\infty dv  \frac {v^{n-1}}{(n-1)!}  \ e^{-l v},
\eea
and simplify the result: 
\bea
&
 	-\frac12\frac {i}{\sqrt{2 \pi}}\int_0^\infty \frac{dl}{l^{1/2}}e^{-l\mu_B} \,\frac{1}{l}
		\int_{w}^\infty dv  \frac{d}{dv}\left[\frac {v^{n}}{n!}  \ e^{-l v}\right]
\nonumber \\  
&\qquad\qquad=
		 \frac {w^{n}}{n!}  \times
		\frac12\frac {i}{\sqrt{2 \pi}}\int_0^\infty \frac{dl}{l^{1/2}}e^{-l\mu_B} \,\frac{1}{l}
		\ e^{-l w} . 
\eea
One may use the string equation  \eqref{MG-hierachy-g=0}
to get 
\be 
\frac{\p^2{\cal F}^c_{(0)}}{\p  \t_0 \t_{n-1}} 
= \frac {w^{n}}{n!}.
\ee 
In addition,   the rest term has the form
\bea
& \frac {i}{\sqrt{2 \pi}}\int_0^\infty \frac{dl}{l^{1/2}}e^{-l\mu_B}
		 \frac{e^{-l w}}{l}
\nonumber \\
&\qquad\qquad	
	=\frac {i}{\sqrt{2 \pi}}\int_0^\infty \frac{dl}{l^{1/2}}e^{-l\mu_B}
		\int_w^\infty dv \,e^{-l v} = - \frac{\partial{\cal F}^o_{(0)}}{\partial \t_{0}}.
\eea
Collecting all the results, we have the open KdV hierarchy \eqref{BMG_open_KdV_0}:
 \be 
  \frac {2n+1}2  
\frac{\partial {\cal F}^o_{(0)}}{\partial \t_n}
 = -\mu_B \frac{\partial{\cal F}^o_{(0)}}{\partial \t_{n-1}}+
 \frac{1}{2}\frac{\partial {\cal F}^o_{(0)}}{\partial \t_0}\frac{\partial^2 {\cal F}^c_{(0)}}{\partial \t_0\partial \t_{n-1}}.
\ee

%%%%%%%%%%%%%%%%%%%%%%%%
\section{Another derivation of generating function on a disk}

Here we demonstrate how \eqref{wave_function} gives the GF of BMG on a disk
for the simplest case \eqref{Fo-muB-space}, 
where all the KdV parameters are turned off $t_{i>0}=0$, except $t_0$.
Extracting the leading term in a series expansion in $\lambda$ of the logarithm of \eqref{wave_function}, 
one finds
\be
\label{wave_function:aapendix}
	F^o_{(0)}(t,z)
	=\sum_{k\geq 0} (t_k-\delta_{k,1}) \frac{z^{2k+1}}{(2k+1)!!}
	- \sum_k  \frac{(2k-1)!!}{z^{2k+1}}\frac{\pd F^c_{(0)}}{\pd t_k}.
\ee
Recalling that
\be 
	{ F}^c_{(0)} = \frac 12 \int^w_0  { P}^2(t,v) \, dv;
	\qquad
	{P}(t, v) = -v+\sum_{m=0}^{\infty} t_m  \frac{v^{m}}{m!},
\ee
one has
\be
	\frac{\pd{F}^c_{(0)}}{\pd t_n} 
	= -\frac{w^{n+2}}{(n+2)\, n!}+\sum_{m=0}^{\infty}  \frac{t_m}{m!\,n!} \frac{ w^{m+n+1}}{(m+n+1)} ,
\ee
whose evaluation at $t_{i>0}=0$ results in
\be
	\frac{\pd{F}^c_{(0)}}{\pd t_n} \bigg|_{t_{i>0}=0}
	=- \frac{t_0^{n+2}}{(n+2)\, n!}+ \frac{ t_0^{n+2}}{(n+1)!} 
	=\frac{t_0^{n+2}}{(n+2)!}.
\ee
Making $t_{i>0}$ turn off, one obtains
\be
	F^o_{(0)}(t_0,z)
	:=
	F^o_{(0)}(t,z)\big|_{t_{i>0}=0}
	=
	t_0z- \frac{z^{3}}{3}
	- \sum_{k=0}^\infty  \frac{(2k-1)!!}{z^{2k+1}}\frac{t_0^{k+2}}{(k+2)!}.
\ee
Substituting $z=i\sqrt{2\mu_B}$ one immediately sees, noting $(-1)!!=1$,
that this is a series expansion of \eqref{Fo-muB-space}:
$F^o_{(0)}(t_0,\mu_B)=i\frac{2\sqrt{2}}{3\lambda}\left(\mu_B+t_0\right)^{3/2}$ for large values of $|\mu_B|$.

%%%%%%%%%%%%%%%%%%%%%%%%
%================================%

\end{document}